# Oscillating paramagnetic Meissner effect and Berezinskii-Kosterlitz-Thouless transition in $Bi_2Sr_2CaCu_2O_{8+\delta}$ monolayer

S. Y. Wang[1#], Y. Yu[1#], J. X. Hao[1], Y. Feng[1], J. J. Zhu[1], Y. S. Lin[1], B. K. Xiang[1], H. Ru[1], Y. P. Pan[1], G. D. Gu[2], K. Watanabe[3], T. Taniguchi[4], Y. Qi[1*], Y. Zhang[1,5*], Y. H. Wang[1,5*]

*1. State Key Laboratory of Surface Physics and Department of Physics, Fudan University, Shanghai 200433, China*
*2. Condensed Matter Physics and Materials Science Department, Brookhaven National Laboratory, Upton, New York 11973, USA*
*3. Research Center for Functional Materials, National Institute for Materials Science, 1-1 Namiki, Tsukuba 305-0044, Japan*
*4. International Center for Materials Nanoarchitectonics, National Institute for Materials Science, 1-1 Namiki, Tsukuba 305-0044, Japan*
*5. Shanghai Research Center for Quantum Sciences, Shanghai 201315, China*

\# These authors contributed equally.

\* To whom correspondence and requests for materials should be addressed. Email: qiyang@fudan.edu.cn; zhyb@fudan.edu.cn; wangyhv@fudan.edu.cn

**Monolayers of a prototypical cuprate high transition-temperature ($T_C$) superconductor Bi$_2$Sr$_2$CaCu$_2$O$_{8+\delta}$ (Bi2212) was recently found to show $T_C$ and other electronic properties similar to those of the bulk[1]. The robustness of superconductivity in an ideal two-dimensional (2D) system was an intriguing fact that defied the Mermin-Wagner theorem[2]. Here, we took advantage of the high sensitivity of scanning SQUID susceptometry[3-7] to image the phase stiffness throughout the phase transition of Bi2212 in the 2D limit. We found susceptibility oscillated with flux between diamagnetism and paramagnetism in a Fraunhofer-like pattern up till $T_C$. The temperature and sample size-dependence of the modulation period agreed well with our Coulomb gas analogy[8,9] of a finite 2D system based on Berezinskii-Kosterlitz-Thouless (BKT) transition[8,10-12]. In the multilayers, the susceptibility oscillation differed in a small temperature regime below $T_C$ in consistent with a dimensional-crossover led by interlayer coupling. Serving as strong evidence of BKT transition in the bulk, there appeared a sharp superfluid density jump at zero-field and paramagnetism at small fields just below $T_C$. These results unified the phase transitions from the monolayer Bi2212 to the bulk as BKT transition with finite interlayer coupling. This elucidating picture favored the pre-formed pairs scenario[10,13,14] for the underdoped cuprates regardless of lattice dimensionality.**

The Mermin-Wagner theorem forbids long-range order at finite temperature in a 2D system with continuous symmetry and short-range interactions. However, a phase transition of infinite order, the famed BKT transition, into a phase without true long-range order is allowed[8,10-12]. In terms of superconducting systems, signs of BKT transitions have been observed in 2D arrays of Josephson junctions[15], ultrathin conventional superconductors[16-19], superconducting interface[20] and YBa$_2$Cu$_3$O$_{7-\delta}$ films[21,22]. Since a BKT transition establishes phase coherence of the preformed paired electrons, similar to Bose-Einstein condensation[13], it is distinct from conventional superconducting phase transition in three dimensions (3D) where pairing and phase coherence happen at the same temperature following the Bardeen-Cooper-Schrieffer (BCS) theory[23]. This is the reason why in conventional 2D superconductors the BKT transition appeared at temperatures below bulk $T_C$.

The question regarding cuprates is whether BKT physics could be universal beyond the 2D limit given their layered structure but finite interlayer coupling[24,25]. Earlier work on bulk crystals showed evidence of vortex excitation above $T_C$[14], supporting a pre-formed pairing scenario[10,13]. Nevertheless, ubiquitous emergent electronic and spin orders are known to be on the same energy scale as Cooper pairing in underdoped cuprates[26-29], which obscured the phase transition between the highly-debated pseudogap regime and the superconducting order[30,31]. Furthermore, the surprise finding of similar electronic structure and $T_C$ in the monolayer and the bulk Bi2212[1] but absence of BKT transitions in both confounds the mystery[32]. Even though the robustness of superconductivity in the monolayer makes it a new platform to investigate the unconventional pairing of cuprates, it first calls for a unifying understanding of the superconducting phase transitions under different lattice dimensions.

Since the BKT transition is fundamentally about vortex-induced phase fluctuation, we look for its signature from the flux rather than charge perspective of the superfluid. The magnetic behavior of a thin superconductor is indeed quite different from a bulk one. Because of the Meissner effect, magnetic field is screened out of a bulk superconductor except around vortices which have a lateral span of London penetration depth $\lambda$ [23]. Diamagnetism is much weaker in thin superconductors when sample thickness $t \ll \lambda$, and therefore magnetic field can go through the entire sample almost unhindered. The characteristic magnetic length scale $\lambda$ in the bulk is replaced by the Pearl length $\Lambda = \frac{2\lambda^2}{t}$, which determines the size of Pearl vortices and Meissner current in thin superconductors[33]. Furthermore, $1/\Lambda$ is directly related to the superfluid density and phase stiffness which are fundamental parameters in BKT physics. (The length scale for the variation of superfluid density, still determined by the coherence length, is not affected by thickness.) The monolayer Bi2212 (half a unit-cell) samples demonstrated recently have provided great opportunities to study the BKT transition of cuprates in a true 2D limit[1,34]. Nevertheless, measuring the weak diamagnetism and Pearl vortices in such exfoliated ultrathin superconductors poses challenges for the sensitivity of magnetometry. Conventional techniques measuring penetration depth such as two-coil inductance and microwave surface impedance lack the spatial resolution and sensitivity for typical exfoliated flakes nanometers in thickness and microns in size.

Scanning superconducting quantum interference device (sSQUID)[3-7] has high flux sensitivity and spatial resolution essential to study Pearl vortices and superfluid density of ultrathin Bi2212. This technique employs a nano-fabricated chip which integrated micron-sized pickup loops into a two-junction SQUID that converted the flux through the loop ($\Phi$) into a voltage signal[35]. The pickup loops of our nano-SQUID were in a gradiometric design so that flux due to uniform external field through both loops cancels out. Therefore, the flux signal we measured was strictly from the sample. In addition, we used mu-metal to shield the earth magnetic field and a home-wound coil to compensate for any residual field so that the sample could be measured in a true zero-field environment. Flowing an alternating current ($I_F$) through the field coil, we obtained the real part of the AC susceptibility ($\chi'$) by demodulating the in-phase component from the flux signal in the pickup loop. We thermally isolated our nano-SQUID from the sample so that the sample temperature could be independently raised up to 200 K without introducing additional noise on the nano-SQUID, which was kept at 4.6 K[7]. As can be seen below, such capability of highly sensitive susceptometry over a large temperature range in a well-controlled magnetic field was critical for the investigation of the phase-transition of cuprate high-temperature superconductors in the 2D limit.

The bulk Bi2212 samples we started from throughout this study were optimally-doped single crystals. In the bulk form, the $T_C$ was 88.2 K as determined by both volumetric magnetometry and sSQUID susceptometry under zero magnetic field (SOM). We mechanically exfoliated the crystals using the technique described earlier[1] to obtain thin flake samples of various thickness (Fig. 1a). The approach curves at 10 K could be well fitted with the model of a thin large diamagnetic disk (Fig. 1b)[36], which gave $\Lambda = 171$ μm for the monolayer (SOM). Using $t = 1.5$ nm of the monolayer, we obtained $\lambda = 358$ nm, comparable to that of the bulk Bi2212 with slight under-doping[37]. The deviation from optimal doping was due to loss of oxygen occurred after exfoliation and before an h-BN capping layer could completely cover it (SOM). This resulted in loss of superconductivity of the left corner of the sample and reduced $T_C$ of the rest comparing with the bulk $T_C$ (Fig. 1c).

Since $\Lambda$ was larger than the size of the monolayer, supercurrent was expected to be mostly along the edge of the sample. Reassuringly, there was no visible vortex in the interior of the monolayer under an out-of-plane field $H = 0.17$ G. The weak magnetic contrast at the edge, which had negative contrast relative to $H$ just inside the boundary (Fig. 1d), suggested it was due to Meissner current. The spatial variation of both the magnetization and susceptibility became much less uniform as temperature increased. Even though χ' (Fig. 1e) was qualitatively similar to its low temperature state, the Meissner current surrounding the whole sample shrank and separated into two loops at 40 K (Fig. 1f). Since the diamagnetism at 40 K was much weaker than that at 15 K (Fig. 1c), meaning a much larger Λ, the smaller and segregated flux feature could only result from formation of domains of different superconducting phases. The sporadic features in magnetometry, also present in the bulk (SOM), were due to vortex moving through the sample during scanning and were intrinsic to Bi2212 at this temperature regime likely due to its strongly layered structure[38]. The domains became more distinctive at 50 K in both χ' and magnetometry images (Figs. 1g and h). The two triangular areas in the middle showed diminishing diamagnetism as they had a lower $T_C = 52$ K (defined by the temperature above which no diamagnetic feature could be observed). These two weakly superconducting domains disrupted the phase coherence between the left and right domains which had $T_C = 64$ K. The existence of such domains could prevent bulk probes from resolving phase-coherent processes within the domains.

At even higher temperatures, susceptometry images changed more dramatically with the field (Fig. 2). The overall diamagnetic signal was weaker at higher fields than those of the lower ones because increased screening current costed free energy and reduced the superfluid density. Magnetometry did not show noticeable Meissner current (Fig. 2a) under a relatively small $H = 0.32$ G at 60 K, but the diamagnetic region shrank and a small area of paramagnetic region occurred on the top (Fig. 2b). At $H = 0.53$ G, the contrast enhanced a bit in magnetometry (Fig. 2c). The paramagnetic region on the top turned into diamagnetic, but another paramagnetic area occurred on the left side of the sample (Fig. 2d). As $H$ increased further to 0.74 G, there was no qualitative change in the magnetometry image (Fig. 2e) but the left side returned to be weakly diamagnetic again with developments of other paramagnetic areas in the middle (Fig. 2f). The magnetometry image at $H = 0.95$ G (Fig. 2g) was similar to the one at $H = 0.74$ G, but the susceptometry was different and the paramagnetic pattern seemed more random (Fig. 2h). The overall diamagnetic signal was weaker at higher fields than those of the lower ones. The weak and similar magnetometry signal ruled out that the paramagnetic signal in susceptometry was from cross-talk between the two channels. The reappearance of diamagnetism at the same location with increasing field suggested that the sample was still in a superconducting state despite the reduced superfluid density under these fields. Such paramagnetism in the superconducting state was reminiscent of the paramagnetic Meissner effect (PME, also called Wohlleben effect), which was observed several decades ago in granular Bi2212 bulk samples by cooling under small magnetic fields[39,40]. The original explanation, which relied on $d$-wave pairing forming $\pi$-junctions across grain boundaries[41], was debated, as the PME may also occur in conventional mesoscopic superconductors[42] and surface states[43] of odd-frequency superconductors[44,45]. Since our Bi2212 sample was single-crystalline, the formation of $\pi$-junctions was unlikely regardless of the pairing symmetry.

In order to investigate the origin of the PME in the monolayer sample, we measured susceptibility as a function of magnetic field and temperature ($T$) at a fixed location on the sample. We picked the middle point of the left region of the monolayer sample (as shown by the black dot in Figure 1g). The susceptibility obtained by sweeping $H$ at various $T$ (Figs. 3a) showed the most pronounced paramagnetism between 50 K and $T_C = 64$ K. Other than the spikes, the field sweep $\chi'$ curves were typical of a type II superconductor: a flat diamagnetic bottom at low field which started to increase at the lower critical field ($H_{c1}$) around 0.5 G (Fig. 3b), which was expected from the penetration depth at this temperature. $\chi'$ leveled off to values slightly below zero at higher fields, which were four orders of magnitude smaller than the upper critical field (> 4.6 $T$ at 61 K for $T_C$ = 64 K). The PME appeared as several symmetric spikes at fields $H_p$ overlaying on the diamagnetic background in the field sweep curves (Fig. 3b), which slowly shifted with temperature (Fig. 3a). The spikes that were closest to zero field (labelled as '1') disappeared below 59 K. The higher order ones (second and third are labelled as '2' and '3', respectively) persisted to lower temperatures. When temperature was continuously varied rather than the field, the *H-T* diagram obtained from the 'field-cooling' cycles showed different peak amplitude but the same peak position in field $H_p$ (SOM). The $H_p$ was also independent of the modulation field frequency or amplitude if it was small, but they were strongly suppressed when the modulation amplitude was comparable or larger than the peak spacing in $H$ (SOM). These ruled out resonance artifact in the AC susceptibility and provided further evidence that the oscillating PME was an intrinsic response of the Bi2212 monolayer under an external field.

The temperature evolution of the PME peaks, in comparison with the diamagnetism at zero-field, showed its connection with BKT transition. The disappearance of the PME peaks was at the same temperature as the disappearance of diamagnetism at zero-field, i.e., $T_C$ = 64 K. The $H_p$ of the lowest three peaks (Fig. 3c) did not go down to zero at $T_C$, suggesting robust phase coherence up to the transition. Converting the zero-field $\chi'(T)$ to $1/\Lambda(T)$ (Fig. 3c, right axis), we found the temperature dependence of the phase stiffness was consistent with that of a BKT transition of a finite 2D superconductor. The interception $1/\Lambda(T_{\text{BKT}}) = \frac{4\pi\mu_0}{\Phi_0^2} k T_{\text{BKT}}$, where $\mu_0$ is the vacuum permeability, $\Phi_0$ the flux quantum and $k$ the Boltzmann constant, determined the BKT temperature $T_{\text{BKT}}$.

In order to investigate the sample size effect on the oscillating PME, we obtained $\chi'(H)$ curves at various horizontal positions $x$ across the left domain (indicated by the arrow in Fig. 1g) of the monolayer sample at 60 K. The lowest three spikes in the *H-T* sweep (Fig. 3a) were still clearly distinguishable at all of the *x* positions except for the one at the left edge (Fig. 3d). All the spikes are shifting towards smaller *H* in a uniform fashion as *x* increased (Fig. 3e). This could be well-understood when we consider field screening from the nano-SQUID (Fig. 1a): sample area on the right side of the pickup loop was completely shielded from the small external field we applied and only the area to the left and underneath the pickup loop was subject to *H*. Normally this would not lead to any observable effect as only the field on the sample directly underneath the pickup loop matters. Here, however, it was the total flux threading the sample rather than the field that determines the susceptibility. As the SQUID moved to larger *x*, the exposed area increased, and a smaller *H* was needed to maintain the same flux. This point could be best shown by multiplying $H_p$ and the exposed area to obtain total flux through the sample as a function of *x* (Fig. 3f). (Since diamagnetism of the monolayer sample at this temperature region was quite weak, as evident from the magnetometry, Meissner screening from the sample could be ignored.) The lowest three peaks respectively corresponded to 1, 2 and 3 flux quanta through the left domain of the sample. The domain size, which was estimated from the area when the nano-SQUID pickup loop was at the right edge of the domain, was about 200 μm², in agreement with the optical image. Using the $H_p(x)$ relation of the three lowest order spikes (Fig. 3e), we can now identify the seemingly random paramagnetic features we observed in the susceptometry images at 60 K (Fig. 2). They were a result of single vortex penetration at *H* = 0.53 G (Fig. 2d), double at *H* = 0.74 G (Fig. 2f) and triple at *H* = 0.95 G (Fig. 2h), respectively. The oscillating PME with *H* was clearly not due to vortex melting[4,46,47], in which case only a singular paramagnetic peak appeared at a large enough field to generate high vortex density for the melting transition[48].

The above observation rather showed that local magnetic susceptibility was dependent upon the global fluxoid of a domain. The scale of the domain size $R \sim 20$ μm over which the coherent oscillation occurred was a thousand times larger than the coherence length of Bi2212 (about 16 nm even at 63 K for $T_C = 64$ K using BCS theory). This ruled out the oscillating PME we observed was the same effect as seen in a conventional mesoscopic superconductor, which required sample size close to its coherence length. By the similar consideration of the disparity in scales, the oscillating PME in monolayer Bi2212 could also not be due to the Little-Parks effect because the change in critical temperature would be unnoticeable: $\frac{\Delta T_C}{T_C} = 0.55(\frac{\xi_0}{R})^2$, where $\xi_0$ is zero-temperature coherence length. As a control experiment supporting the above two arguments, ultrathin NbSe$_2$ flakes, which has very similar $\xi_0$ and penetration depth as monolayer Bi2212, did not show any PME over a similar normalized temperature ($T/T_C$) and field range (SOM). Indeed, vortex cores have diverging size through a conventional superconducting phase transition which suppresses phase coherence.

The persistence of phase coherence even at $T_C$, manifested by the susceptibility oscillation, was a unique feature of a BKT transition, which required finite-sized phase singularities. Previous theoretical and numerical studies of a 2D XY model under weak frustration did find paramagnetic susceptibility[49]. But even before any modeling, the duality between phase and charge of a 2D superconductor allowed us to visualize vortices as Coulomb gas[8,9] to gain deeper intuition for the oscillation of PME. The $\ln(r/\xi)$ repulsion between two vortices with the same vorticity distance $r$ apart is exactly the same as Coulomb repulsion in 2D electrodynamics. The magnetic field acts as a chemical potential for the Coulomb gas and thus the vorticity is equivalent to the vortex particle number. When this chemical potential exceeds the energy cost of a vortex, one vortex will be thermally excited. Moreover, adding an additional vortex to our sample of finite size must overcome the effective repulsion between the two vortices, which originates from redistribution of supercurrents when two vortices are present. Consequently, the second vortex will be excited only when the chemical potential reaches the amount proportional to the inter-vortex repulsion. Further increasing the magnetic field, the third vortex will be excited when the chemical potential overcomes the mutual repulsion between the three vortices. Therefore, upon increasing the magnetic field, vortices will enter one-by-one at specific fields. This is in direct analogy with the Coulomb blockade of single-electron transistors[50,51] that would exhibit spikes in tunneling conductance when gate voltage was tuned. Noting that the flux counterpart of the charge tunneling conductance is magnetic susceptibility, we can see that the observed oscillating PME spikes originated from the quantized nature of vortex.

Mapping the 2D superconductor near the BKT transition to the Coulomb-gas model, we computed the paramagnetic response of the vortices in a circular disk. The high symmetry of the geometry simplified the computation although admittedly it did not match the actual shape of the sample. We found qualitative agreement on $\chi'(T, H)$ between the modeling (with only two fitting parameters) and the experiment (SOM). The extracted first two PME peaks evolving with temperature also agreed with those of the experiment (Fig. 3c, blue and green). The main discrepancy occurred for the third peak where the modeling underestimated the peaks (Fig. 3c). This is likely caused by our crude approximation of the shape of the sample. Nevertheless, the modeling captured the most striking feature of the experiment: the $H_p$'s were finite at $T_C$ (Fig. 3c). Indeed, in the Coulomb-gas model, the peaks remain at finite fields because the superfluid density renormalizes both the repulsive interaction and the chemical potential of the vortices

(SOM). This could only happen if there existed well-defined vortex cores and finite coherence length much smaller than sample size, which were essential for a vortex-driven phase transition.

Besides the monolayer, the oscillating PME occurred generally in the ultrathin Bi2212 samples we studied. For example, a quadruple-layer sample ($T_C = 87$ K) similarly exhibited paramagnetic peaks (Fig. 4a). The main difference from the monolayer was that the peaks at $H = \pm 0.2$ G did not persist beyond 85 K, at which temperature the phase stiffness also exhibited a kink (Fig. 4b). The PME of a different domain on this sample with slightly lower $T_C$ (SOM) and a quintuple-layer sample were similarly separated into two temperature regimes (Figs. 4c and d). The absence of oscillating PME in the temperature regime above the kink suggested it was not caused by disparate $T_C$'s of the surface layer and the buried ones. Instead, for a layered system with small interlayer coupling, the BKT transition *should* have two characteristic temperatures determined by $1/\Lambda(T_{BKT}^n) = \frac{16\pi\mu_0}{\Phi_0^2}\frac{t}{d\cdot n}kT_{BKT}^n$, where $d$ is the thickness of the monolayer and $n$ is the number of phase-independent layers[52,53]. $n = 1$ corresponded to a lower transition where all the layers are Josephson-coupled and $n = t/d$ for a higher one where all the layers were independent. The quadruple-layer showed these two $T_{BKT}^n$ for $n = 1$ and 4 below and above the kink, respectively (Fig. 4b). This strongly suggested the kink was a cross-over regime for the onset of interlayer coupling. The PME with much enhanced oscillation below the cross-over (Figs. 4a-d) further corroborated that the phase coherence between the layers was established in this regime.

In even thicker samples, inhomogeneous domains in different layers were more prominent such that their contributions to PME were less in phase. This is clearly the case in a 20-nm sample where the PME peaks occurred at seemingly random fields (SOM). In a bulk sample, which was more than 200 nm thick, the domains were even more 'coarse-grained'. As a result, the PME peaks did not show any oscillation with field and only a small range of temperature close to $T_C$ exhibited overall paramagnetic response (Fig. 4e). As a function of temperature, there was a PME peak right below $T_C$ at finite fields (Fig. 4f, red), reminiscent of the original observation of PME in cuprates by bulk magnetometry[39,40]. The striking similarity suggested they shared the same origin (even though those observations were on poly-crystalline samples). The PME temperature-range grew with $H$ because of the reduction of diamagnetic strength with field. Such PME was clearly absent from the bulk NbSe$_2$ ($T_C$ = 6.6 K) control sample (SOM). Noise-like features in susceptibility of Bi2212 below $T_C$ (Fig. 4e), which was absent above $T_C$, suggested strong phase-fluctuations accompanying vortex excitations in this temperature regime.

The phase stiffness versus temperature curves of Bi2212 and NbSe$_2$ were also markedly different (Fig. 4f). While the latter showed a smooth power-law rise below $T_C$ typical of a BCS superconductor, Bi2212 showed a sharp jump which rose to 85% of its peak magnitude within 0.15% of $T_C$. Such a sharp jump in superfluid density at zero-field was exactly what a BKT transition entailed[54]. At finite fields, the PME peak slightly obscured the sharp rise in diamagnetic susceptibility. Notwithstanding, the PME *per se* was strong evidence that BKT physics was also responsible for the superconducting to normal transition of the bulk Bi2212. The BKT transition we observed in Bi2212 is likely applicable to cuprate superconductors in general since the family shares the layered structure with various degrees of interlayer coupling. Cooper pairs formed at a very high temperature enables vortex excitation to play a central role behind the universal scaling between the superfluid density and $T_C$[13,55], the abnormal critical behavior of the specific heat of underdoped YBa$_2$Cu$_3$O$_{6+x}$[30] and the Nernst effect in the pseudo-gap phase[14].

In conclusion, we have observed PME oscillations in monolayer and few-layer Bi2212, which evolved to be continuous with field in thick samples. Combined with the characteristic features in phase stiffness at precisely-determined zero-field, we provided strong evidence that the superconducting transitions in underdoped Bi2212 from 2D to the bulk were generalized BKT transitions with interlayer coupling. The spatial resolution of sSQUID to distinguish domains of different size and $T_C$ and its sensitivity in susceptometry at very low field and high temperature were indispensable to these observations. Our technique and the observations showed that ultrathin Bi2212 was a promising material platform to understand phase fluctuations in the enigmatic pseudogap regime of the underdoped cuprates.

**Acknowledgement**

Y.H.W. would like to acknowledge partial support by the Ministry of Science and Technology of China under Grant No. 2017YFA0303000, NSFC Grant No. 11827805 and Shanghai Municipal Science and Technology Major Project Grant No. 2019SHZDZX01. K.W. and T.T. acknowledge support from the Elemental Strategy Initiative conducted by the MEXT, Japan (Grant Number JPMXP0112101001), JSPS, KAKENHI (Grant Numbers JP19H05790 and JP20H00354) and A3 Foresight by JSPS. The work at BNL was supported by the US Department of Energy, office of Basic Energy Sciences, contract No. DOE-sc0012704. YQ acknowledges financial support from NSFC Grant No. 11874115. Y.Y. and Y.Z. acknowledges financial support from National Key Research Program of China (Grant No. 2018YFA0305600), NSF of China (Grant Nos. U1732274 and 11527805), Shanghai Municipal Science and Technology Commission (Grant No. 2019SHZDZX01) and Strategic Priority Research Program of Chinese Academy of Sciences (Grant No. XDB30000000). Y.Y. also acknowledges support from the China Postdoctoral Science Foundation (Grant Nos. BX20180076 and 2018M641907). All the authors are grateful for the experimental assistance by X. H. Chen and D. L. Feng and for the stimulating discussions with Y. Chen, C. Varma, Ziqiang Wang, T. Xiang and Q. H. Wang.

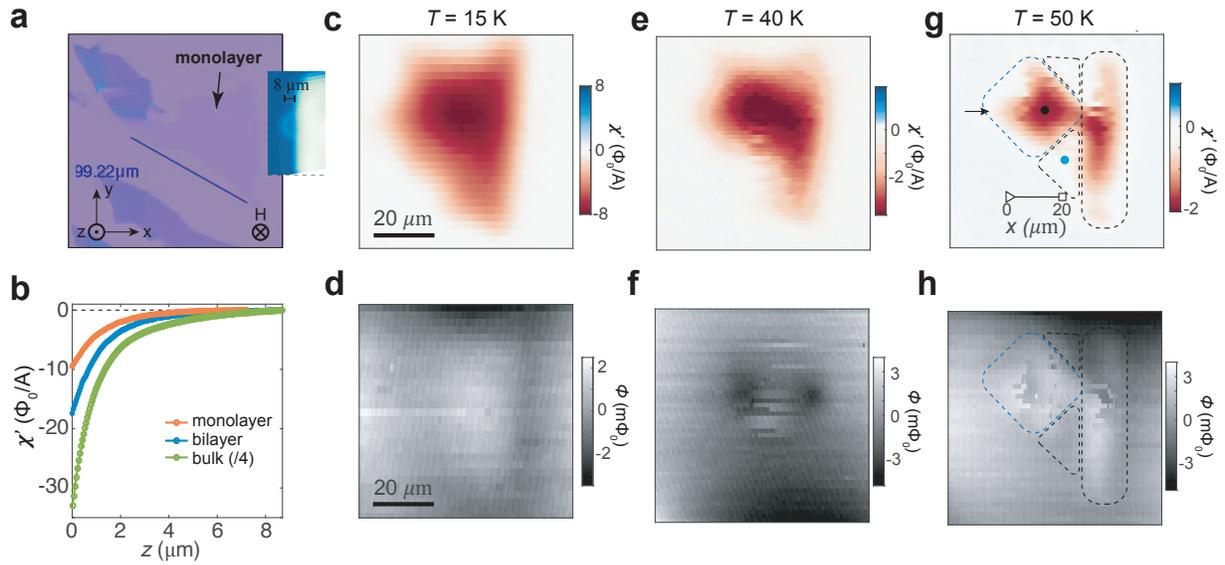

**Fig. 1 sSQUID magnetometry and susceptometry of an underdoped Bi2212 monolayer at various temperatures. a,** Optical image of the monolayer sample. Inset: susceptometry image of a nano-SQUID head showing the orientation of the field coil and the screening layer made of Nb in the same scale as the optical image of the monolayer sample. Only the area of the sample to the left of the screening layer was subject to an out-of-plane field $H$. **b,** Susceptibility approach curves of the monolayer (orange), bilayer (blue) and bulk (green) samples. Diamagnetism corresponds to $\chi' < 0$. **c** and **d,** Susceptometry and magnetometry images of the monolayer at 15 K, respectively, with $H = 0.17$ G. **e** and **f,** and **g** and **h**, Corresponding images at 40 K and 50 K, respectively. The discontinuous jump in magnetometry was a result of sporadic vortex motion intrinsic to the vortex glass state at this temperature and field regime. The four dashed boxes in **g** and **h** roughly outlined four domains. The two triangular domains had lower $T_C = 52$ K and the diamond and the rectangular ones $T_C = 64$ K.

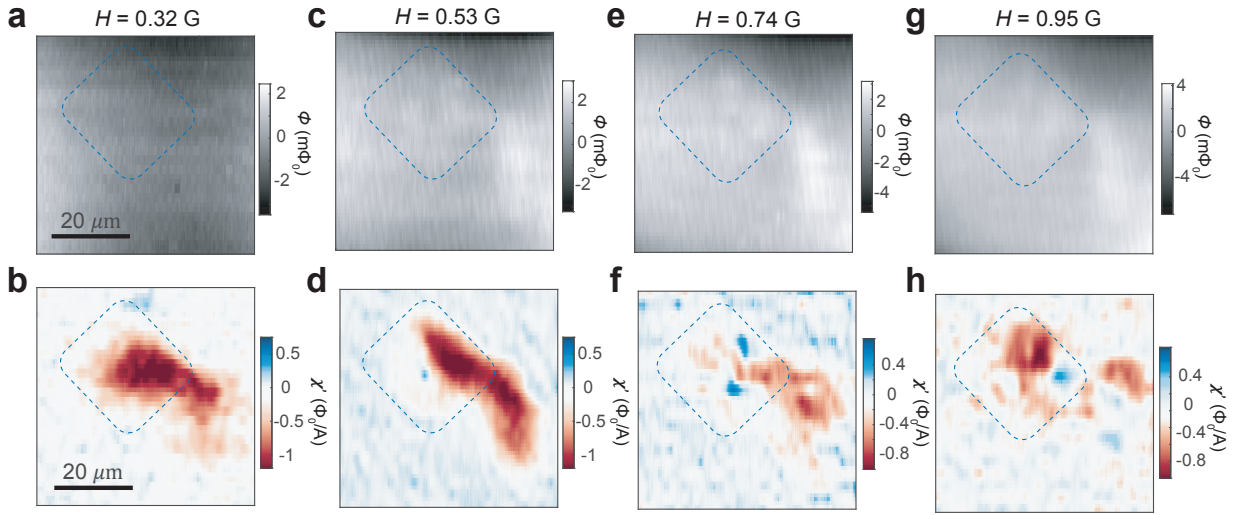

**Fig. 2 Appearance of paramagnetic regions under a small external magnetic field at elevated temperatures. a-h,** Magnetometry and susceptometry images for the monolayer sample shown in **Fig. 1** obtained at 60 K and various external field $H$ as indicted on the panels. The dashed boxes delineate the region of the diamond-shaped domain in **Fig. 1g**. Paramagnetic Meissner effect appeared as blue patches in susceptometry images which returned back to diamagnetic with increasing field.

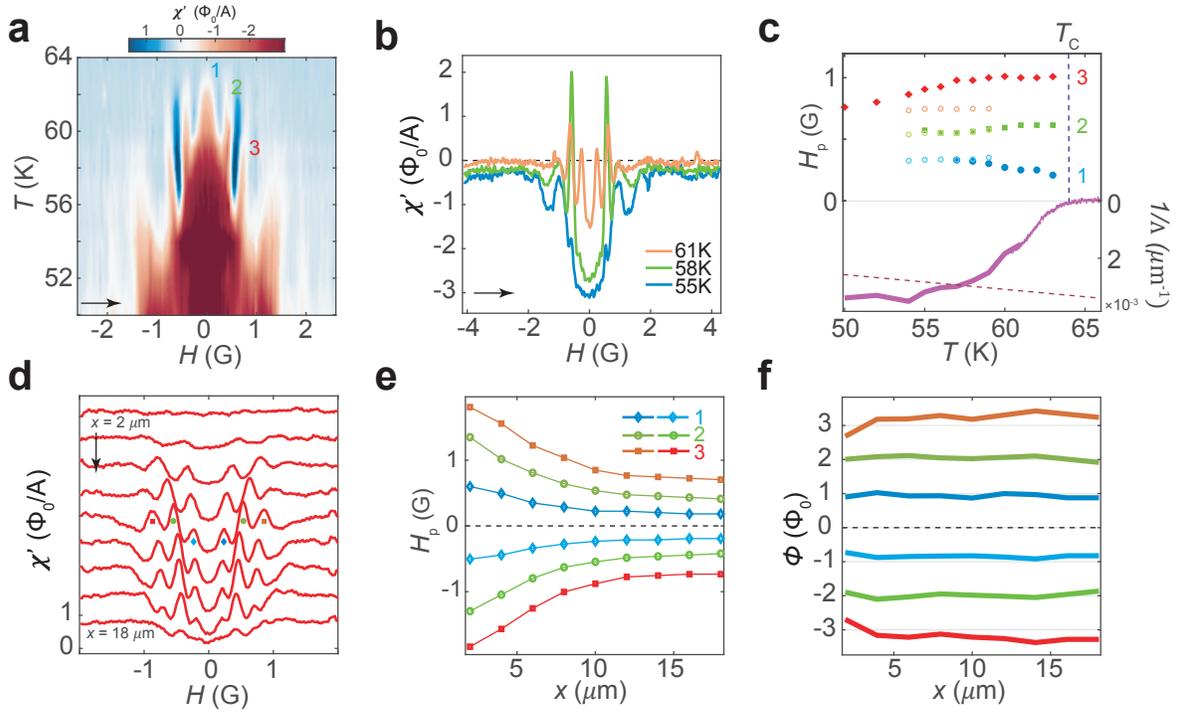

**Fig. 3 Oscillating paramagnetic susceptibility in the monolayer as a function of external field.** The data was obtained on the diamond-shaped domain in **Fig. 1g**. **a,** $\chi'(T, H)$ of the monolayer sample taken at the point shown in **Fig. 1g**. The data were obtained by sweeping the field at different $T$. Arrows indict the direction of the sweep. The three lowest order paramagnetic peaks were labeled by the numbers '1', '2', and '3'. **b,** $\chi'(H)$ at various temperatures taken from **a**. **c,** Peak position ($H_p$) as a function of $T$ extracted from **a** (left axis). The three lowest-order paramagnetic peaks were represented by blue, green and red d symbols, respectively. The light open circles of similar colors were obtained from our Coulomb gas simulation (see text and SOM). The inverse of the Pearl length, $1/\Lambda(T)$, obtained from the diamagnetic susceptibility at zero-field in **a** (right axis). The diamagnetic data in the range of 60 K – 66 K (solid dots) were from a fine temperature sweep at zero-field (SOM). The interception of the dashed straight line with $1/\Lambda(T)$ determined the BKT temperature (see text). **d,** $\chi'(H)$ curves obtained at 60 K at different displacement $x$ of the nano-SQUID tip over the monolayer sample along the arrow shown in **Fig. 1g**. The curves with larger $x$ were shifted downward proportionally. **e,** $H_p(x)$ extracted from **d**. **f,** $H_p(x)$ from **e** plotted in units of flux quantum. This was obtained by integrating the field over the exposed area of the domain (see **Fig. 1a** and text).

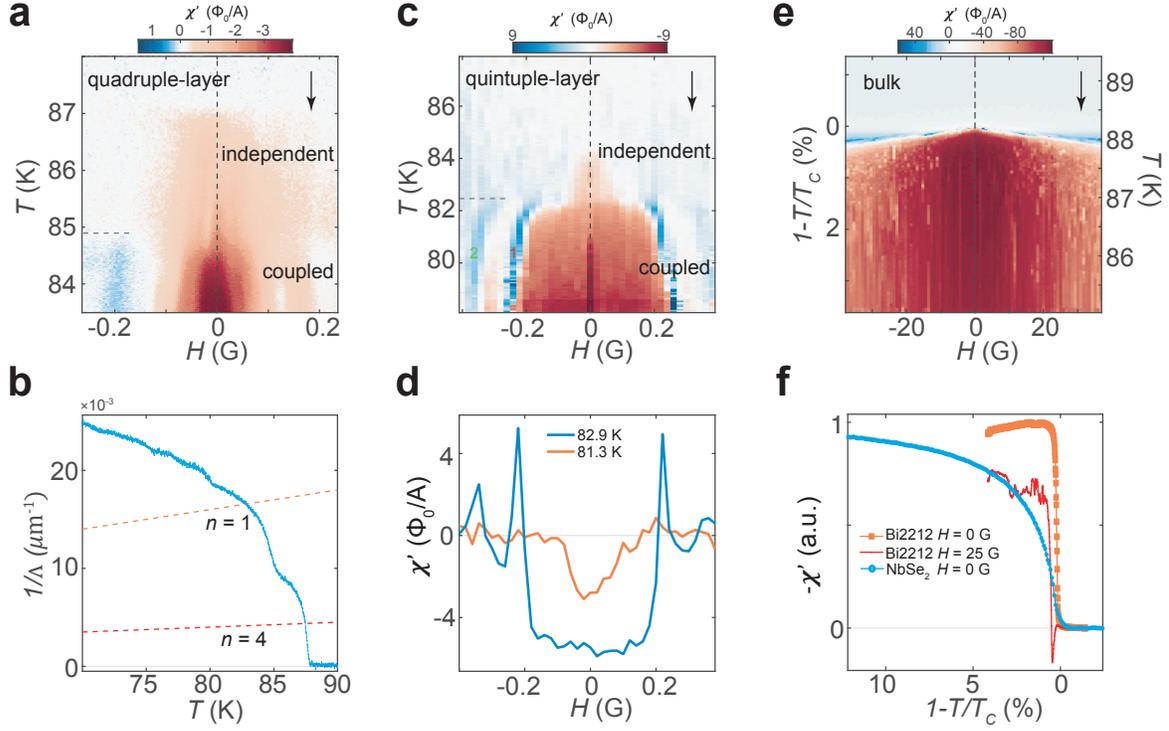

**Fig. 4 Paramagnetic Meissner effect in multilayers and a bulk sample. a,** $\chi'(T, H)$ in a quadruple-layer sample. **b,** $1/\Lambda(T)$ of the quadruple-layer sample obtained from $\chi'(T)$ at zero-field. The interception of the dashed lines with this curve determines the BKT temperature of the coupled layers ($n = 1$) and independent layers ($n = 4$), respectively. **c,** $\chi'(T, H)$ in a quintuple-layer sample. **d,** $\chi'(H)$ at two different temperatures obtained from **c**. The oscillating paramagnetic susceptibility in the multilayers separated into two temperature regimes demarked by the horizontal dashed line, below which the layers were Josephson-coupled. **f,** $\chi'(T)$ at zero-field (orange squares) and $H = 25$ G (red dots) taken from **e**. In comparison, a similar zero-field curve for a NbSe$_2$ bulk sample ($T_C = 6.6$ K) was shown as the blue curve. Note that the 'noisy' diamagnetic signal in the Bi2212 bulk sample under finite $H$ was absent above $T_C$ and was therefore intrinsic to the sample. Both the paramagnetic peak at finite $H$ and a sharp jump of superfluid density at $T_C$ suggested BKT transition in the bulk Bi2212. Additional data on these samples available in SOM.

Content of the Supplementary Information